\documentclass[aps,nofootinbib,prl,superscriptaddress,tightenlines,
notitlepage,twocolumn,floatfix]{revtex4-1}
\usepackage{tabularx}
\usepackage{graphicx}
\usepackage{epsfig}
\usepackage{amssymb,bm,tensor}
\usepackage{textcomp}
\usepackage{amsmath}
\usepackage[varg]{txfonts}
\usepackage{enumerate}
\usepackage{mathtools}
\usepackage[normalem]{ulem}
\usepackage{bm} 
\usepackage[OT1]{fontenc}
\usepackage[colorlinks]{hyperref}
\usepackage{multirow}
\usepackage{makecell}
\usepackage{soul}
\usepackage{booktabs}
\usepackage{nameref}
\usepackage{threeparttable}
\usepackage{xspace}

\usepackage[capitalize]{cleveref}
\newcommand{\sacrak}{\texttt{SACRA-K}\xspace}

\usepackage[dvipsnames]{xcolor}
\usepackage{orcidlink}
\hypersetup{colorlinks=true,
            citecolor=NavyBlue,
            linkcolor=NavyBlue,
            urlcolor=NavyBlue}

\begin{document}
\title{Subsolar-mass binary mergers of strange stars and neutron stars: 
gravitational waves and ejecta}

\date{\today}

\author{Yong Gao\,\orcidlink{0000-0003-1390-5477}}
\email{yong.gao@aei.mpg.de}
\affiliation{Max Planck Institute for Gravitational Physics (Albert Einstein Institute), 14476 Potsdam, Germany}

\author{Ming-Zhe Han\,\orcidlink{0000-0001-9034-0866}}
\affiliation{Max Planck Institute for Gravitational Physics (Albert Einstein Institute), 14476 Potsdam, Germany}
\affiliation{Key Laboratory of Dark Matter and Space Astronomy, Purple Mountain Observatory, Chinese Academy of Sciences, Nanjing, 210033, People's Republic of China}

\author{Kenta Kiuchi\,\orcidlink{0000-0003-4988-1438}}
\affiliation{Max Planck Institute for Gravitational Physics (Albert Einstein Institute), 14476 Potsdam, Germany}
\affiliation{Center of Gravitational Physics and Quantum Information, Yukawa Institute for Theoretical Physics, Kyoto University, Kyoto, 606-8502, Japan}

\author{Masaru Shibata\,\orcidlink{0000-0002-4979-5671}}
\affiliation{Max Planck Institute for Gravitational Physics (Albert Einstein Institute), 14476 Potsdam, Germany}
\affiliation{Center of Gravitational Physics and Quantum Information, Yukawa Institute for Theoretical Physics, Kyoto University, Kyoto, 606-8502, Japan} 

\author{Enping Zhou\,\orcidlink{0000-0002-9624-3749}}
\affiliation{School of Physics, Huazhong University of Science and Technology, Wuhan 430074, China}

\author{Kenta Hotokezaka\,\orcidlink{0000-0002-2502-3730}}
\affiliation{Research Center for the Early Universe, Graduate School of Science, The University of Tokyo, Bunkyo, Tokyo 113-0033, Japan}
\affiliation{Max Planck Institute for Gravitational Physics (Albert Einstein Institute), 14476 Potsdam, Germany}

\begin{abstract}
We present the first numerical-relativity simulations of subsolar-mass binary
strange star (SS) mergers and compare with binary neutron star (NS) mergers across
equations of state, masses, and mass ratios. The self-bound nature of SSs makes
them less deformed during the inspiral and keeps a sharp surface up to contact,
driving strong shock heating and a large radial bounce that are far weaker in the
NS. The more compact SS thus reaches a higher gravitational-wave cutoff frequency
$f_\mathrm{cut}$ before contact but a lower post-merger peak frequency $f_2$. Within each class
these frequencies follow quasi-universal relations with the tidal deformability,
and their ratio $f_2/f_\mathrm{cut}$ cleanly separates the two classes. Both
classes can eject $\sim10^{-2}\,M_\odot$ of material, neutron-rich for the NS and
decompressed quark matter for the SS, a potential source of an electromagnetic
counterpart whose observation could test the SS and NS hypotheses for subsolar-mass events.
\end{abstract}

\maketitle

\emph{Introduction}---The LIGO--Virgo--KAGRA (LVK) network has now catalogued over 300 compact-binary
coalescences~\cite{LIGOScientific:2026wfs,LIGOScientific:2025slb}. The vast majority are binary black
holes (BBHs), with two binary neutron star (BNS)
events~\cite{LIGOScientific:2017vwq,LIGOScientific:2017ync} and three confident
neutron-star--black-hole binaries~\cite{LIGOScientific:2021qlt,LIGOScientific:2024elc}. 
Every confidently measured
component mass lies above $1\,M_\odot$~\cite{LIGOScientific:2021djp}, consistent
with standard stellar evolution. A growing effort is under way to probe the
\emph{subsolar} regime. Dedicated pipelines have scanned LVK data from the third
and fourth observing runs for subsolar-mass binaries~\cite{Nitz:2022ltl,Kacanja:2026byy,LIGOScientific:2026wxz},
and candidate events have surfaced---GW231109\_235456~\cite{Niu:2025nha} and the
O4 alerts S250818k and S251112cm~\cite{GraceDB:S250818k,GraceDB:S251112cm}---though
none has yet withstood scrutiny as a confident detection. A candidate optical
counterpart to S250818k was even identified~\cite{Kasliwal:2025keb}, but follow-up
favoured a Type~IIb supernova over a kilonova~\cite{Hall:2025qsm,Ackley:2026zlf}.

A subsolar event would point to new physics, such as a primordial black hole~\cite{Zeldovich:1967,Carr:1974nx,Sasaki:2018dmp,Baumgarte:2026azn,Haque:2026yum,Crescimbeni:2024cwh,Miller:2024rca}
or an unconventionally formed NS, because standard stellar evolution cannot produce a subsolar NS: both core-collapse supernovae and the accretion-induced collapse of
white dwarfs are set by the Chandrasekhar mass, which yields a minimum NS mass of
$\sim\!1.17\,M_\odot$~\cite{Suwa:2018uni,Muller:2024aod}, consistent with the
lightest measured NS~\cite{Martinez:2015mya} (though see Ref.~\cite{Tauris:2019sho}). 
Hints of lighter objects nonetheless
exist: the mass of the central compact object in the supernova remnant HESS\,J1731$-$347 is 
inferred to be $0.77^{+0.22}_{-0.17}\,M_\odot$~\cite{Doroshenko:2022nwp}.

Subsolar NSs might instead form in gravitationally unstable collapsar disks, where 
the low electron fraction $Y_e$ of neutron-rich fragments suppresses the
local Chandrasekhar mass and allows collapse at subsolar
mass~\cite{Piro:2006ja,Metzger:2024ujc,Chen:2025wbz,Wu:2026hth}, although for this scenario, 
an unusually efficient cooling process would be necessary 
(see also a discussion in~\cite{Nakamura:1989pc} for a relevant topic).
Alternatively, a subsolar compact star could be a strange star (SS). If strange
quark matter---deconfined $u$, $d$, and $s$ quarks---is the true ground state of
baryonic matter, as conjectured by Bodmer and
Witten~\cite{Bodmer:1971we,Witten:1984rs,Farhi:1984qu,Alcock:1986hz}, then SSs are
self-bound and not subject to a Chandrasekhar-type collapse threshold, and so extend naturally to
subsolar masses. Such low-mass SSs have
been proposed to form as relics of the early
Universe~\cite{Witten:1984rs,Shao:2025tec}, through the collapse of rapidly
rotating NSs~\cite{Nakamura:2002qt}, or via the accretion-induced collapse of a
white dwarf whose core is converted to strange matter~\cite{Xu:2004sf,DiClemente:2022ktz}.

\begin{figure*}[t]
  \centering
  \includegraphics[width=0.95\linewidth]{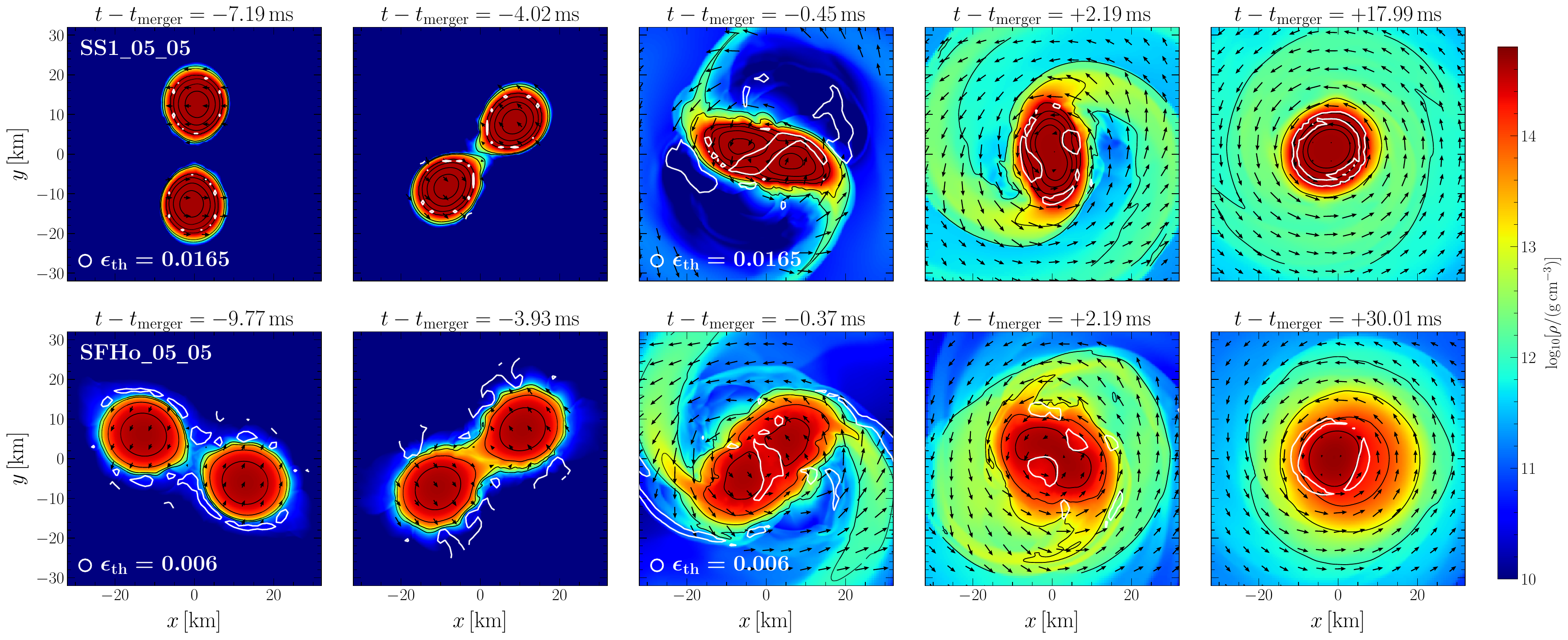}
  \caption{Equatorial-plane rest-mass density
    $\log_{10}[\rho/(\mathrm{g\,cm^{-3}})]$ (colour) for the equal-mass
    $0.5{+}0.5\,M_\odot$ SS (SS1, top row) and NS (SFHo, bottom row)
    binaries, shown at five times relative to merger $t-t_\mathrm{merger}$
    (indicated above each panel), from the late inspiral (left) to the
    post-merger remnant (right).  Black curves are rest-mass-density
    iso-contours at $\log_{10}[\rho/(\mathrm{g\,cm^{-3}})]=12,\,13,\,14.57,\,
    14.59,\,14.61,\,14.63$ for the SS and $12,\,13,\,14,\,14.5$ for the NS.  
    White curves are iso-contours of the
    thermal part of the specific internal energy, $\epsilon_\mathrm{th}$, at the
    value labelled in the first panel of each row ($0.0165$ for the SS, $0.006$
    for the NS), tracing the shock-heated contact layer.  Arrows show the
    in-plane coordinate velocity.  The compact,
    self-bound SS is only weakly deformed during inspiral and drives a strong
    shock and a large radial bounce at contact, whereas the larger, more tidally
    deformable NS develops pronounced spiral arms and sheds mass before
    merger; both relax to a rotating remnant surrounded by an extended disk. 
    The animations of the snapshots and corresponding GWs can be found at \cite{subsolar}.}
  \label{fig:snapshot}
\end{figure*}

Unlike BHs, NSs and SSs are tidally deformed during the inspiral~\cite{Hinderer:2007mb,Damour:2009vw}
and can eject matter at merger to power a kilonova~\cite{Li:1998bw,Metzger:2010sy} (see, e.g.,~\cite{Shibata:2019wef,Metzger:2019zeh}
and references therein). The tidal deformation accelerates the
inspiral and leaves an imprint on the gravitational-wave (GW) signal, encoded at leading order by the
mass-weighted dimensionless tidal deformability $\tilde\Lambda$~\cite{Flanagan:2007ix,Favata:2013rwa,Wade:2014vqa,Read:2013zra,Hotokezaka:2016bzh,Hotokezaka:2013mm}.
This signature strengthens steeply toward subsolar masses, with the deformability
of an individual star growing from $\mathcal{O}(10^3)$ at $1\,M_\odot$ to
$\mathcal{O}(10^6$--$10^7)$ at $0.1\,M_\odot$, making the subsolar regime
exceptionally sensitive to the equation of state (EOS)~\cite{Hinderer:2009ca,Silva:2016myw,Gao:2021uus}. Search templates that
neglect these matter effects may lose a significant fraction of their sensitive
volume~\cite{Bandopadhyay:2022tbi}, while the same imprint already lets current
detectors separate a subsolar NS from a BBH~\cite{Corman:2026lbt,Crescimbeni:2024cwh}.

The subsolar regime is, moreover, a promising place to tell a subsolar SS from an
NS. An SS is self-bound and compact, whereas an NS is gravitationally bound and
extended, and the two evolve in opposite directions toward lower masses. The NS
grows more extended, reaching radii $R\sim11$--$14\,\mathrm{km}$ at $M \sim 0.5\,M_\odot$ and
larger still at lower masses, whereas the self-bound SS shrinks as $R\propto
M^{1/3}$ to $\sim\!8\,\mathrm{km}$ and below. The SS is thus
systematically less deformable, and the two classes follow $\Lambda$--$M$ relations
of markedly different slope (\cref{fig:eos}). In principle the inspiral tidal deformability can
separate them, but only for very loud signals or by combining many
events~\cite{Crescimbeni:2024qrq,Wang:2024xon}. A soft-EOS NS and an SS also overlap in
$\tilde\Lambda$ across a wide range of masses and EOSs, leaving the leading-order
tidal signature degenerate. The merger itself, however, is essentially unexplored
in the subsolar regime. We therefore ask how subsolar SS and NS binaries merge---what
governs their dynamics, their characteristic merger and post-merger
gravitational-wave frequencies, and their mass ejection---and whether these break
the inspiral degeneracy between the two classes.

Such simulations remain scarce in the subsolar regime. On the NS side, numerical relativity (NR) has
so far treated a binary in which only one of the two components is a subsolar star~\cite{Corman:2026lbt}, whereas binary SS
mergers have been modelled only at canonical or higher masses, either with
smoothed-particle hydrodynamics under the conformal-flatness
approximation~\cite{Bauswein:2009im} or in full general
relativity~\cite{Zhou:2021tgo,Zhu:2021xlu,Grippa:2024ppo}. The
self-bound SS surface is a sharp, near-discontinuous density drop that standard
hydrodynamics resolves only with care~\cite{Zhou:2021upu,Chen:2023bxx}.
In this Letter we present NR simulations of subsolar-mass binary SS
mergers for the first time and systematically contrast their dynamics, gravitational-wave spectra, and
ejecta with those of subsolar binary NS mergers. Throughout, we use geometric units with
$G=c=1$, where $G$ and $c$ are the gravitational constant and the speed of light, respectively.

\emph{Numerical models and grid setup}---Each class spans a stiff-to-soft
range of EOSs---DD2, SFHo, and WFF1 for the NSs and the (quasi-)MIT-bag family
SS1--SS3~\cite{Zhang:2020jmb} for the SSs (\cref{fig:eos})---across component
masses $0.3$--$0.7\,M_\odot$ and mass ratios $q=1$, $1.22$, and $1.5$
(\cref{tab:models}). We construct the quasi-equilibrium initial data with
\texttt{FUKA}~\cite{Papenfort:2021hod}, augmented with a self-bound surface solver
for the SSs. The binaries start at a separation of $\approx33$--$37\,\mathrm{km}$, with the
orbital eccentricity iteratively reduced to $\lesssim7\times10^{-4}$.

We evolve the fully general-relativistic hydrodynamics with \sacrak~\cite{Han:SACRAK}, a GPU-accelerated port of
\texttt{SACRA-MPI}~\cite{Yamamoto:2008js,Kiuchi:2017pte,Kiuchi:2022ubj}, in the
moving-puncture Baumgarte--Shapiro--Shibata--Nakamura formulation~\cite{Shibata:1995we,Baumgarte:1998te,Baker:2005vv,Campanelli:2005dd} with the Z4c
constraint-propagation prescription~\cite{Hilditch:2012fp} and fourth-order finite differencing. The hydrodynamics uses a high-resolution
shock-capturing scheme---the Harten--Lax--van~Leer-contact (HLLC) Riemann solver (cf.~the \texttt{NANASI} code~\cite{Kiuchi:2022ubj,Kiuchi:2026pgb})
for the NSs and the Harten--Lax--van~Leer (HLLE) solver for the
SSs, whose sharp self-bound surface requires a more numerically robust scheme. Shock heating
is captured by a $\Gamma$-law thermal component added to each cold EOS, with
$\Gamma_\mathrm{th}=4/3$ for the SSs and $1.75$ for the NSs. The grid is a
box-in-box mesh of ten $2{:}1$ refinement levels, the finest four comoving with
each star ($2N\times2N\times N$ points, equatorial symmetry), evolved at three
resolutions (\cref{tab:models}). The code's parallel setup and performance are
summarized in \cref{tab:performance}.
We extract GWs at $r\approx700\,\mathrm{km}$
and identify the dynamical ejecta with the Bernoulli criterion $-hu_t>1$,
with $h$ the specific enthalpy and $u_t$ the time component of the four-velocity.

\begin{figure}[t]
  \centering
  \includegraphics[width=\columnwidth]{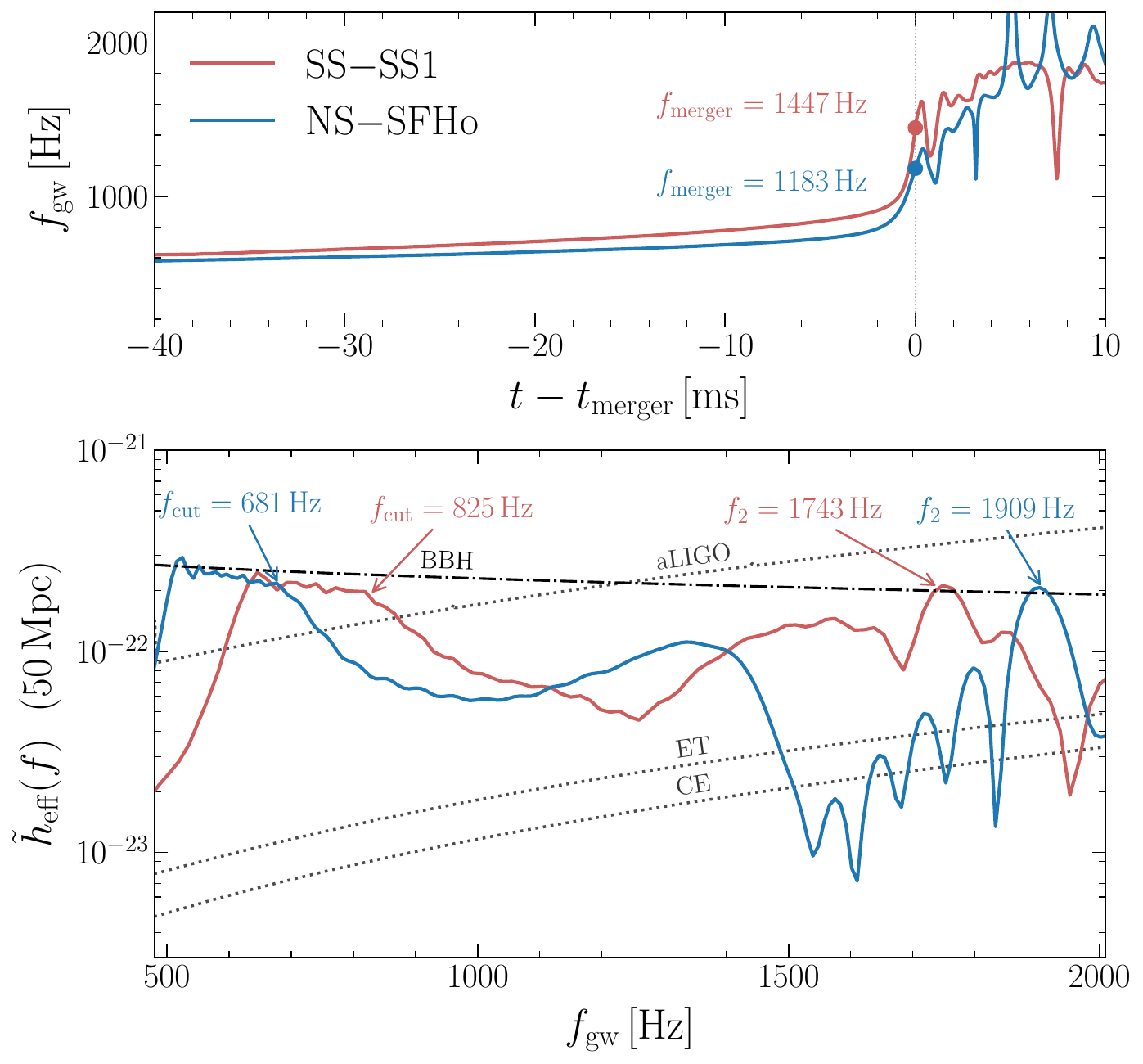}
  \caption{GW signal from the representative equal-mass
    $0.5+0.5\,M_\odot$ NS--SFHo (blue) and SS1 (red) binaries.
    \emph{Top}: instantaneous frequency $f_\mathrm{gw}$ of the $\ell=m=2$ mode
    versus retarded time relative to merger.
    The filled circles mark the value at the amplitude peak.
    \emph{Bottom}: effective amplitude $\tilde{h}_\mathrm{eff}(f)$ for a source at
    $50\,\mathrm{Mpc}$, marking the cutoff frequency $f_\mathrm{cut}$ and the
    post-merger peak $f_2$, with representative detector noise curves (aLIGO,
    Einstein Telescope, Cosmic Explorer) and a nonspinning binary-black-hole
    inspiral (BBH) for reference.}
  \label{fig:gw_time_freq}
\end{figure}

\emph{Merger dynamics and GW signal.}---An SS is self-bound by the strong
interaction and of nearly uniform density, whereas an NS is gravitationally
bound, centrally condensed, and more extended.  This structural difference shapes
how the two objects merge, which we illustrate with representative equal-mass
$0.5{+}0.5\,M_\odot$ SS and NS binaries.  The SS (SS1) is more compact:
$R\simeq8.4\,\mathrm{km}$, $C\equiv M/R\simeq0.088$, and dimensionless tidal deformability
$\Lambda=5.2\times10^4$, versus $12.4\,\mathrm{km}$, $0.060$, and
$8.2\times10^4$, respectively, for the NS (SFHo).  We follow their merger through the
rest-mass-density snapshots of \cref{fig:snapshot} and the GW
signal of \cref{fig:gw_time_freq}: the instantaneous GW frequency as a function of retarded time (\emph{top}) and the effective
spectrum $\tilde{h}_\mathrm{eff}(f)$ for a face-on source at a hypothetical distance of $D=50\,\mathrm{Mpc}$ (\emph{bottom}). 

With the same chirp mass, the two binaries follow the point-particle inspiral
almost identically at large separation, where the effective GW amplitude scales
approximately as $\tilde{h}_\mathrm{eff}\propto f^{-1/6}$.  As tidal effects become sizeable
in the late inspiral, $\tilde{h}_\mathrm{eff}$ departs from this nonspinning-BBH
baseline and steepens; the break defines the cutoff frequency $f_\mathrm{cut}$
(lower panel of \cref{fig:gw_time_freq}), which we locate by fitting a broken
power law to $\tilde{h}_\mathrm{eff}(f)$.  

The first column of \cref{fig:snapshot}
shows the density field at $f_\mathrm{cut}$. Below this frequency the binary follows a quasi-circular
inspiral and radiates strongly; above it tidal effects accelerate the late inspiral
off this track and the binary evolves rapidly toward merger, so its GW frequency
sweeps up and $\tilde{h}_\mathrm{eff}$ drops steeply~\cite{Kiuchi:2010ze}.
The more compact SS has smaller tidal deformability, so its cutoff lies
higher, $f_\mathrm{cut}\simeq825\,\mathrm{Hz}$ against $681\,\mathrm{Hz}$ for the
NS. Beyond $f_\mathrm{cut}$ the two binaries reach merger by different mechanisms.
As shown in the second column of \cref{fig:snapshot}, the more compact SS keeps its binary-like
structure up to contact and becomes dynamically unstable once the binary reaches
the innermost stable circular orbit~\cite{Limousin:2004vc,Gondek-Rosinska:2008zmv,Friedman:2001pf}.
The more extended NS instead sheds mass through its surface before contact, where
its orbital kinetic energy is dissipated.

\begin{figure}[t]
  \centering
  \includegraphics[width=0.95\columnwidth]{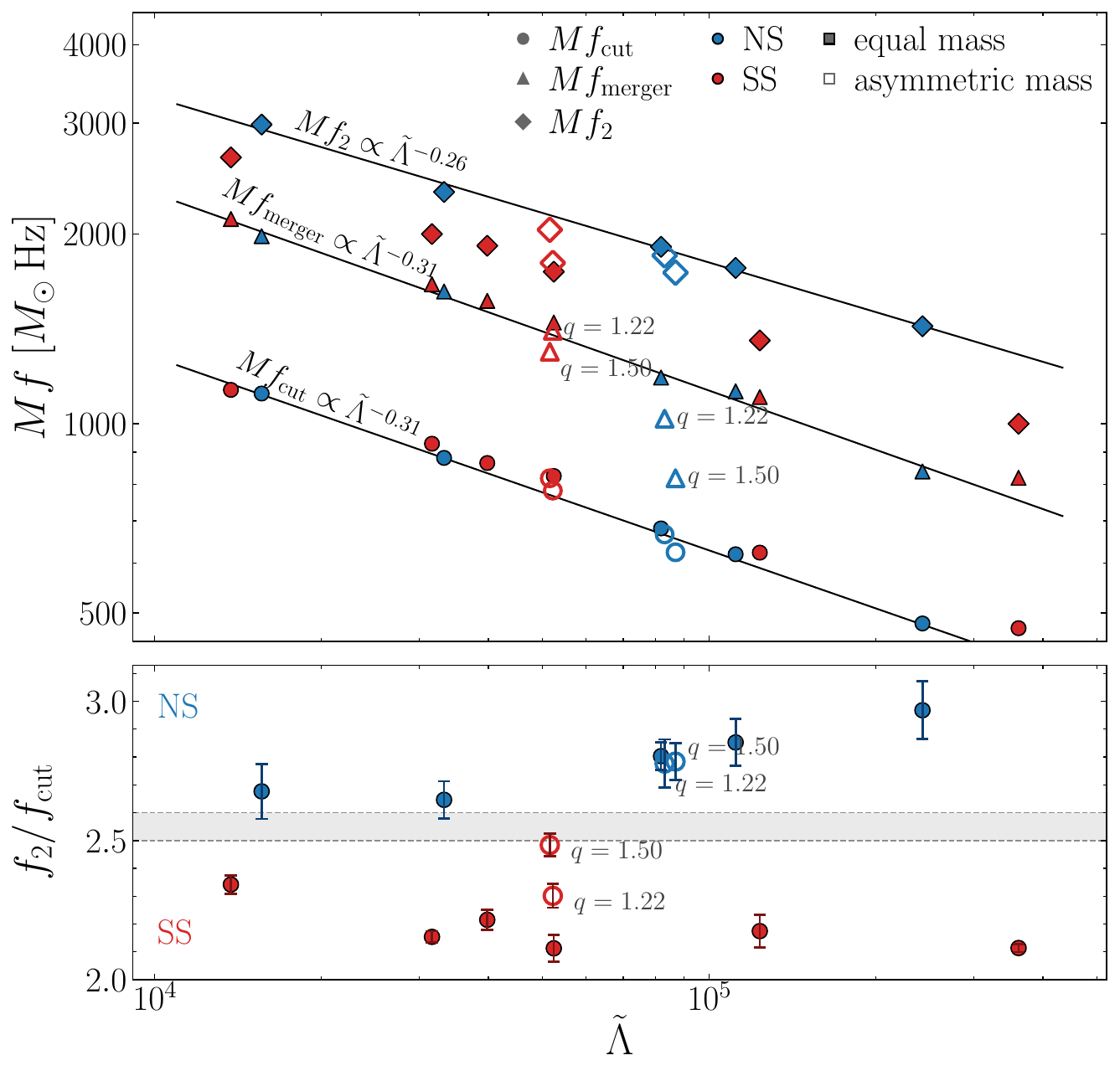}
  \caption{Characteristic GW frequencies of all binaries in the grid, scaled by
    the total mass $M$ and plotted against the mass-weighted tidal
    deformability $\tilde\Lambda$.  \emph{Top}: cutoff frequency
    ($M f_\mathrm{cut}$, circles), frequency at the amplitude peak
    ($M f_\mathrm{merger}$, triangles), and dominant post-merger peak
    ($Mf_2$, diamonds) for NS (blue) and SS (red); filled symbols are
    equal-mass, open symbols unequal-mass, with the mass ratio $q$ marked next
    to the corresponding open triangles.  Solid
    lines are power-law fits to the NS sequence (all NSs for $f_\mathrm{cut}$, 
    the equal-mass NSs only for $f_\mathrm{merger}$ and $f_2$). \emph{Bottom}: the ratio
    $f_2/f_\mathrm{cut}$, which cleanly separates the two classes; the shaded band
    at $f_2/f_\mathrm{cut}=2.5$--$2.6$ marks the gap between the SS branch (below)
    and the NS branch (above). Markers use the
    finest-resolution value. The error bar is the half-spread $(\max-\min)/2$ across the
    three grid resolutions evolved per binary (\cref{tab:models}).}
  \label{fig:correlation}
\end{figure}

Because the SS has a smaller tidal deformability and reaches merger at a smaller separation, its GW amplitude peaks at
a higher frequency, $f_\mathrm{merger}=1447\,\mathrm{Hz}$ against
$1183\,\mathrm{Hz}$ for the NS.  The merger shown in the third and fourth columns
of \cref{fig:snapshot} is far more violent for the SS. Its self-bound cores
collide at higher velocity, and the steep pressure gradient at the contact surface
drives a strong shock that heats the layer to a peak thermal energy
$\epsilon_\mathrm{th}\gtrsim0.03$, hotter and more sharply confined than in the NS
(white $\epsilon_\mathrm{th}$ contours at $0.0165$ for the SS against $0.006$ for the
NS).  The shocked double cores then rebound from a compressed contact at $\sim7\,\mathrm{km}$ out to
$\simeq14\,\mathrm{km}$ (third column).  This large radial bounce raises spiral arms that shed mass
at $t-t_\mathrm{m}\sim1$--$2\,\mathrm{ms}$ (fourth column).  The NS contact is
milder.  Being more extended and having a larger tidal deformability, it instead develops pronounced
spiral arms already before merger at $t-t_\mathrm{m}\sim-2\,\mathrm{ms}$ to $-1\,\mathrm{ms}$.

The post-merger remnants then oscillate at the dominant post-merger frequency
$f_2$ for tens of milliseconds and settle into differentially rotating
configurations surrounded by extended disks. The angular velocity rises 
from a slowly rotating core to a maximum at intermediate
radii and then declines outward as in the canonical-mass NS mergers. As shown in the lower panel of \cref{fig:gw_time_freq},
$f_2$ is lower for the SS, $\simeq1750\,\mathrm{Hz}$ against
$\simeq1900\,\mathrm{Hz}$ for the NS, even though before merger the SS is more
compact.  This anticorrelation originates from the SS's
large radial bounce and shock re-expansion at merger, which lower the average density of the
oscillating remnant.

\emph{Quasi-universal relations and discrimination.}---The characteristic
frequencies described above differ between the two classes, but each also depends on the
EOS, mass, and mass ratio, so a single frequency difference is not by itself
diagnostic. To disentangle these dependences we span a grid of three EOSs per
class, component masses $0.3$--$0.7\,M_\odot$, and mass ratios $q=1$, $1.22$, and
$1.5$. We then seek quasi-universal relations linking the characteristic GW
frequencies $f_\mathrm{cut}$, $f_\mathrm{merger}$, and $f_2$ to the mass-weighted
dimensionless tidal deformability,
\begin{equation}
  \tilde\Lambda=\frac{16}{13}\,
  \frac{(m_1+12m_2)\,m_1^4\,\Lambda_1+(m_2+12m_1)\,m_2^4\,\Lambda_2}{(m_1+m_2)^5},
  \label{eq:lamtilde}
\end{equation}
the leading-order tidal imprint on the waveform. For canonical-mass NS binaries,
$f_\mathrm{merger}$ and $f_2$ are known to obey such relations with $\tilde\Lambda$,
and a departure of the $f_2$--$\tilde\Lambda$ relation has been proposed as a
signature of a strong phase transitions in the remnant~\cite{Bauswein:2018bma,Weih:2019xvw,Most:2018eaw,Lam:2024azd}. 
We test whether
these relations persist in the subsolar regime and how the self-bound SSs depart
from them.

\cref{fig:correlation} collects the three frequencies, each scaled by the total
mass $M$, against $\tilde\Lambda$. We evolved three resolutions per binary
(\cref{tab:models}) and use the finest-resolution value of each frequency in the fits below. The
half-spread $(\max-\min)/2$ across resolutions is $\lesssim2.6\%$, $\lesssim5\%$, and $\lesssim3.6\%$
for $f_\mathrm{cut}$, $f_\mathrm{merger}$, and $f_2$, respectively, and $\lesssim3.7\%$ for the ratio
$f_2/f_\mathrm{cut}$ shown as error bars in the lower panel, well below the inter-class differences. We fit $f_\mathrm{cut}$ to all binaries but $f_\mathrm{merger}$ and
$f_2$ to the equal-mass binaries only, because in the asymmetric systems the strong tidal force alters the merger
dynamics and shifts these frequencies (an effect also found for canonical-mass BNS~\cite{Kiuchi:2019kzt}), as we discuss separately.
Each follows a power law $M f=A\,\tilde\Lambda^{p}$
within its class. Writing the pair $(A,p)$ with $A$ in units of
$10^4\,M_\odot\,\mathrm{Hz}$, the NSs give $(2.2,-0.307)$, $(4.1,-0.313)$, and
$(3.7,-0.262)$ for $f_\mathrm{cut}$, $f_\mathrm{merger}$, and $f_2$, and the SSs
give $(1.5,-0.272)$, $(3.4,-0.292)$, and $(4.4,-0.296)$. The exponents agree
between the classes to within $\sim0.04$ but with different amplitudes.  The
residual scatter about the fits is at most $2.2\%$.

At fixed $\tilde\Lambda$ the equal-mass SS binaries sit up to $\sim12\%$ above the NS
$f_\mathrm{cut}$ relation, well beyond the $2.2\%$ residual scatter---a genuine class
difference set by the self-bound nature and smaller radii of SSs. Their $f_2$, however, lies $\sim13$--$22\%$ below,
a much larger and oppositely signed shift. This shift is dynamical rather than
structural. The violent radial bounce and shock re-expansion at merger lower the
average density of the oscillating remnant, so $f_2$ falls even though the SS is
the more compact star before merger. These offsets largely persist at unequal
mass.
The cutoff frequency, fixed during the inspiral, stays on the relations
at any mass ratio. The merger frequency, by contrast, is sensitive to the tidal
disruption of the lighter star at contact. For the extended NSs it falls
$\sim15\%$ below the relation at $q=1.22$ and $\sim31\%$ at $q=1.5$, whereas the
more compact SSs deviate by only $\sim2$--$10\%$. The post-merger $f_2$ likewise
separates the classes, except for the $q=1.5$ SS, where the strong tidal field
disrupts the companion at merger and the radial bounce is weak, so $f_2$ rises onto
the NS relation---an extreme and isolated case.

Because $f_\mathrm{cut}$ is offset upward and $f_2$ downward for SSs, the two shifts
compound in the ratio $f_2/f_\mathrm{cut}$ (lower panel of \cref{fig:correlation}).
It clusters at $2.11$--$2.34$ for the equal-mass and $q=1.22$ SSs and at
$2.65$--$2.97$ for the NSs. Even the extreme $q=1.5$ SS, at $2.48$, stays below the
NS minimum, so the two classes do not overlap across the numerical models. We therefore expect
$f_2/f_\mathrm{cut}$ to be a robust discriminant between self-bound and gravitationally bound subsolar stars.
For that marginal $q=1.5$ SS case, the merger frequency provides an independent check. Because the
more compact SS resists tidal disruption until close to contact, its $f_\mathrm{merger}$ departs from
the universal relation by only $\simeq10\%$, whereas the extended NS, disrupted well before contact,
falls $\simeq31\%$ below it.

\begin{figure}[t]
  \centering
  \includegraphics[width=\columnwidth]{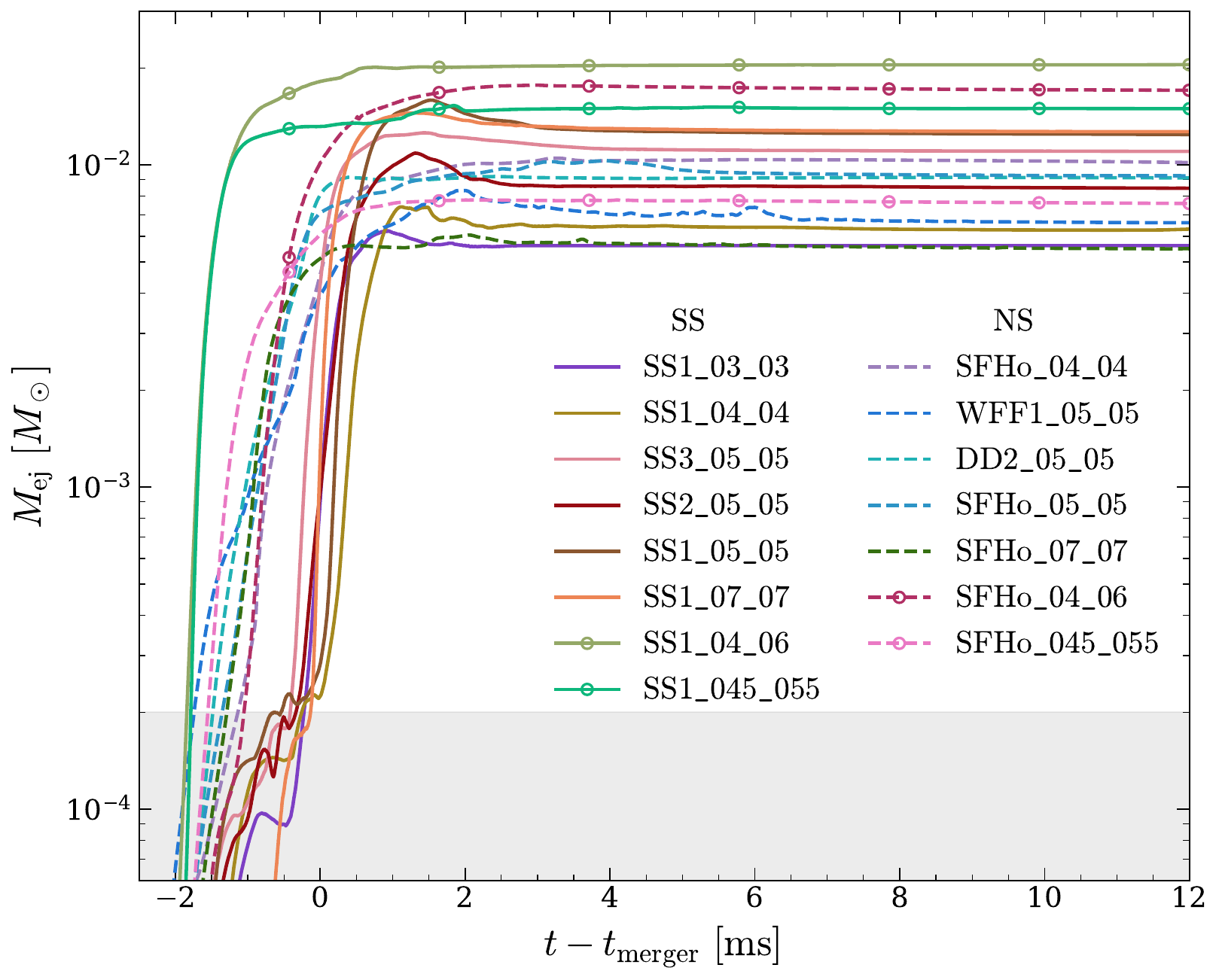}
  \caption{Ejecta mass $M_\mathrm{ej}$ as a function of time
    relative to merger, for the SS (solid) and NS
    (dashed) binaries, computed from the Bernoulli criterion ($-hu_t>1$). Open circles mark the unequal-mass runs ($q=1.22$ and $1.50$).
    Each curve uses the highest-resolution run of the binary.
    The shaded band ($M_\mathrm{ej}<2\times10^{-4}\,M_\odot$) marks the level
    of the baryonic mass-conservation error, below which the measured ejecta is
    not reliable. This upper bound is reached only by the unequal-mass SS binaries,
    while for all other runs the conservation error stays below
    $\sim10^{-5}\,M_\odot$.
    }
  \label{fig:ejecta}
\end{figure}

\emph{Ejecta.}---In both classes the ejecta mainly originates in the tidal tail raised
around merger, but the mechanism that unbinds it differs. \cref{fig:ejecta} shows
the dynamical ejecta rest mass $M_\mathrm{ej}$, measured with the Bernoulli
criterion $-hu_t>1$ (see \cref{tab:models} for the geodesic criterion $-u_t>1$),
for all the models. Because the NS is more extended and has a larger tidal deformability,
it sheds matter through spiral arms that become unbound already
before merger. 
The far more compact SS unbinds matter at merger, when the violent shock re-expansion of
the merged core raises spiral arms and injects energy and angular momentum into them, producing the
steplike, post-merger rise of $M_\mathrm{ej}$. Most configurations settle to
$M_\mathrm{ej}\sim10^{-2}\,M_\odot$ within $\sim10\,\mathrm{ms}$ of merger. The average velocity of the dynamical ejecta is in the range of 0.12--0.17$c$.

Since the unbinding mechanism differs, the two classes respond oppositely as
the binary is made lighter and more deformable. Along the SFHo equal-mass
sequence, $M_\mathrm{ej}$ rises from $5.5\times10^{-3}\,M_\odot$ at a
total mass of $1.4\,M_\odot$ to $1.0\times10^{-2}\,M_\odot$ at
$0.8\,M_\odot$, as the lighter NS becomes more deformable and sheds a stronger
tidal tail. Along the SS1 sequence it instead falls, from
$1.3\times10^{-2}\,M_\odot$ at $1.4\,M_\odot$ to
$6.3\times10^{-3}\,M_\odot$ at $0.8\,M_\odot$ and
$5.6\times10^{-3}\,M_\odot$ at $0.6\,M_\odot$, as the lighter,
less compact SS drives a weaker collision and weaker shock re-expansion,
tempering the mass ejection. In
the unequal-mass binaries of both classes ($q=1.22$ and $1.5$) the heavier
companion raises a massive tidal tail that begins ejecting before merger
and dominates the ejecta budget, reaching the largest masses
in the grid, up to $\sim2\times10^{-2}\,M_\odot$. The differentially
rotating remnant and its disk can unbind a further, substantial
amount of mass through disk outflows on longer
timescales~\cite{Metzger:2014ila,Fujibayashi:2017puw,Shibata:2019wef,Siegel:2017nub, Fujibayashi:2020dvr,Kiuchi:2022nin,Kiuchi:2023obe}, 
adding to the total ejecta budget.

The subsolar BNS ejecta is expected to be neutron-rich, powering a bright
kilonova (e.g.,~\cite{Kawaguchi:2019nju, Fujibayashi:2020dvr}) whose non-detection would argue against a subsolar BNS. The SS ejecta is instead decompressed strange quark matter
that may fragment into hot strangelets. Whether these evaporate into neutron-rich
nucleons, driving an $r$-process and a kilonova much as the NS does, or survive as
bound, electromagnetically dark quark nuggets is mainly set by the binding energy of
quark matter at zero pressure and by the ejecta
temperature~\cite{Alcock:1985vc,Madsen:1998uh,Bucciantini:2019ivq,Miao:2024qik}.

The merger leaves a rapidly rotating remnant with a rotational energy of
$\sim10^{52}\,$erg, and it likely acquires a large magnetic
field~\cite{Price:2006fi,Kiuchi:2015sga,Kiuchi:2023obe,Kiuchi:2026pgb}. Such a field would
spin the remnant down through magnetic-dipole radiation, injecting a sizable
fraction of this energy into the surrounding ejecta. The energized ejecta then
drives a strong shock into the interstellar medium, producing synchrotron emission
from radio to X-ray
wavelengths~\cite{Metzger:2013cha,Hotokezaka:2015eja,Horesh:2016dah}. In addition to GWs and a kilonova, this transient would provide
a further means to test the existence of subsolar-mass SS and NS mergers.

\emph{Acknowledgments}---We thank the members of the Computational Relativistic 
Astrophysics group in AEI for helpful discussions.  
Numerical computations were performed on the clusters Sakura, Raven, and Viper at the Max Planck 
Computing and Data Facility. This work was in part supported by 
Grant-in-Aid for Scientific Research (grant Nos. 23H04900, 23H01172 23H01169, and 23K25869) of 
Japanese MEXT/JSPS, the JST FOREST Program (JPMJFR2136), the National Natural Science Foundation of China under Grants No.~12233011, 
the Project for Young Scientists in Basic Research (Grant No.~YSBR-088) of the Chinese Academy of Sciences.

\bibliography{refs}

\clearpage
\section{Supplemental material}

\setcounter{section}{0}
\setcounter{equation}{0}
\setcounter{figure}{0}
\setcounter{table}{0}
\renewcommand{\thesection}{S\arabic{section}}
\renewcommand{\theequation}{S\arabic{equation}}
\renewcommand{\thefigure}{S\arabic{figure}}
\renewcommand{\thetable}{S\arabic{table}}

\subsection{Equations of state and equilibrium properties}
\label{app:eos}

\begin{figure}[t]
  \centering
  \includegraphics[width=\columnwidth]{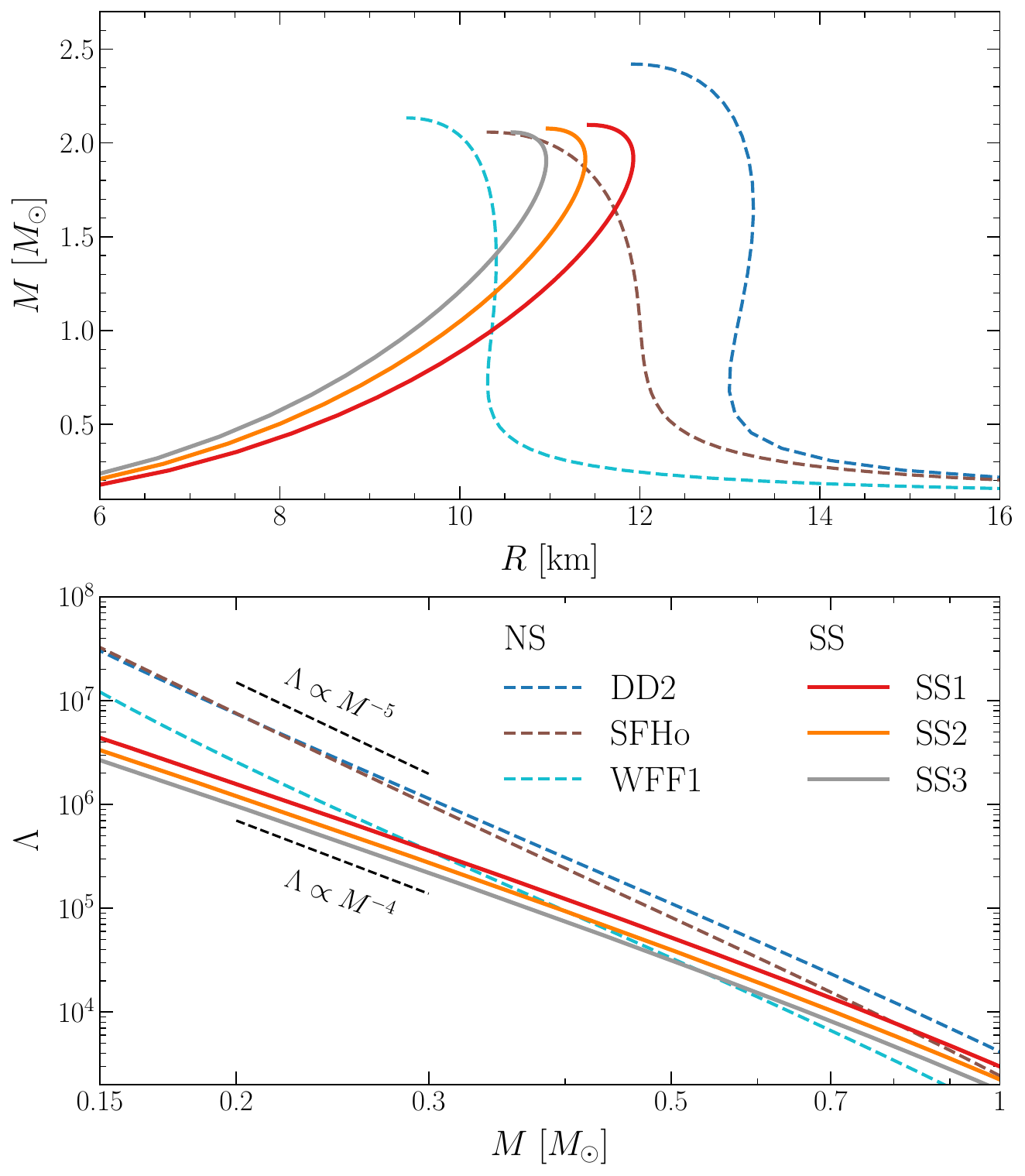}
  \caption{Mass--radius relation (top) and dimensionless tidal deformability
    $\Lambda$ versus mass (bottom) for the SS EOSs SS1, SS2, SS3
    (solid) and the NS EOSs SFHo, WFF1, DD2 (dashed), shown down to
    $0.15\,M_\odot$.  At a fixed subsolar mass the SSs are more compact
    than the NSs, with correspondingly smaller tidal deformabilities.}
  \label{fig:eos}
\end{figure}

For the SSs we adopt the unified interacting quark-matter EOS of the modified
MIT bag model~\cite{Zhang:2020jmb}, which describes cold, charge-neutral,
$\beta$-equilibrated three-flavor quark matter through
\begin{equation}
  p=\frac{1}{3}\left(\epsilon-4B\right)
   +\frac{4\lambda^2}{9\pi^2}\left[\sqrt{\,1+\frac{3\pi^2(\epsilon-B)}{\lambda^2}}\,-1\right],
  \label{eq:bag}
\end{equation}
where $\epsilon$ is the energy density, $B$ the effective bag constant, and
$\lambda^2$ a coupling parameter that absorbs the leading perturbative-QCD
correction, the strange-quark mass, and color-superconducting
pairing~\cite{Zhang:2020jmb,Zhou:2017pha,Alford:2004pf}. The pair $(B,\lambda^2)$ fixes the EOS,
and the matter is self-bound: $\epsilon$ stays finite as $p\to0$, so each star ends
in a sharp surface whose density is set by $B$. We evolve three models, SS1, SS2,
and SS3, with $(B,\lambda^2)=(52.4,\,0)$, $(75.0,\,38.9)$, and
$(96.0,\,157.3)\,\mathrm{MeV\,fm^{-3}}$. Larger $B$ raises the surface density and
larger $\lambda^2$ stiffens the EOS, so the stars become more compact from SS1 to
SS3, with smaller tidal deformability at fixed mass. For the NSs we use three
nuclear EOSs spanning the range from stiff to soft: DD2 (stiff), SFHo, and WFF1 (soft).

\cref{fig:eos} shows the resulting stellar models, whose tidal
deformabilities separate the two classes most clearly through the slope of
$\Lambda(M)$. Writing $\Lambda=\tfrac{2}{3}k_2\,C^{-5}$ with $k_2$ the
quadrupole Love number and $C=M/R$ the compactness, the logarithmic slope
decomposes as
\begin{equation}
  \frac{d\ln\Lambda}{d\ln M}
   =5\,\frac{d\ln R}{d\ln M}-5+\frac{d\ln k_2}{d\ln M}.
  \label{eq:slope}
\end{equation}
The $-5$ from $\Lambda\propto C^{-5}$ is common to both classes, so the radius
term is what distinguishes them. At low mass a self-bound SS behaves as a nearly
incompressible drop with $R\propto M^{1/3}$ (\cref{fig:eos}), contributing $+5/3$.
With $d\ln k_2/d\ln M\approx-0.5$ this gives $d\ln\Lambda/d\ln M\approx-3.9$ over
$0.3$--$0.7\,M_\odot$, almost identical for SS1--SS3. 
A gravitationally bound NS
instead keeps a slightly growing radius toward lower mass
($d\ln R/d\ln M\approx-0.05$ to $-0.1$), so the radius term nearly cancels and the
slope steepens to $\approx-4.6$ (DD2) and $-4.9$ (SFHo).

\clearpage
\onecolumngrid
\subsection{Initial parameters and merger outputs}
\label{app:models}

\begin{table}[!ht]
\caption{Initial parameters and merger outputs for all numerical models.
  For each model we list the component masses ($m_1{+}m_2$), circumferential radii ($R_1,R_2$), and
  mass-weighted tidal deformability $\tilde\Lambda$ (\cref{eq:lamtilde}),
  followed by the results at up to three
  grid resolutions, labelled by the finest grid spacing $\Delta x_\mathrm{finest}$.
  The cutoff, peak, and post-merger frequencies
  $f_\mathrm{cut},f_\mathrm{merger},f_2$ are defined in the main text. The dynamical
  ejecta rest mass $M_\mathrm{ej}$ and mass-averaged velocity
  $\langle v\rangle_\mathrm{ej}=\sqrt{2T/M_\mathrm{ej}}$, with kinetic energy $T=E-U-M_\mathrm{ej}$
  ($E$ and $U$ the total and internal energy of the unbound matter), are evaluated
  $10\,\mathrm{ms}$ after merger, with both the geodesic criterion ($-u_t>1$) and
  the Bernoulli criterion ($-hu_t>1$).}
\label{tab:models}
\footnotesize
\setlength{\tabcolsep}{4.6pt}
\begin{tabular}{l c c c c c c c c c c c}
\toprule
 & & & & & & & & \multicolumn{2}{c}{Geodesic} & \multicolumn{2}{c}{Bernoulli} \\
\cmidrule(lr){9-10}\cmidrule(lr){11-12}
Model & $m_1{+}m_2$ $[M_\odot]$ & $R_1,R_2$ [km] & $\tilde\Lambda$ & $\Delta x_{\rm finest}\,[\rm m]$ & $f_\mathrm{cut}$ [Hz] & $f_\mathrm{merger}$ [Hz] & $f_2$ [Hz] & $M_\mathrm{ej}$ $[10^{-2}M_\odot]$ & $\langle v\rangle_\mathrm{ej}$ $[c]$ & $M_\mathrm{ej}$ $[10^{-2}M_\odot]$ & $\langle v\rangle_\mathrm{ej}$ $[c]$ \\
\midrule
\multirow{3}{*}{SS1} & \multirow{3}{*}{$0.3{+}0.3$} & \multirow{3}{*}{$7.1$} & \multirow{3}{*}{$3.6{\times}10^5$} & $184.6$ & $798.3$ & $1331.5$ & $1666.4$ & $0.07$ & $0.111$ & $0.15$ & $0.094$ \\
 & & & & $138.4$ & $798.3$ & $1350.9$ & $1666.6$ & $0.39$ & $0.123$ & $0.43$ & $0.120$ \\
 & & & & $110.7$ & $787.7$ & $1365.8$ & $1665.1$ & $0.40$ & $0.132$ & $0.56$ & $0.130$ \\
\addlinespace
\multirow{3}{*}{SS1} & \multirow{3}{*}{$0.4{+}0.4$} & \multirow{3}{*}{$7.8$} & \multirow{3}{*}{$1.2{\times}10^5$} & $197.0$
 & $780.6$ & $1356.8$ & $1749.8$ & $0.30$ & $0.121$ & $0.36$ & $0.116$ \\
 & & & & $148.0$ & $777.8$ & $1379.5$ & $1785.0$ & $0.12$ & $0.118$ & $0.15$ & $0.111$ \\
 & & & & $118.1$ & $779.0$ & $1376.8$ & $1694.0$ & $0.54$ & $0.128$ & $0.63$ & $0.123$ \\
\addlinespace
\multirow{3}{*}{SS1} & \multirow{3}{*}{$0.45{+}0.55$} & \multirow{3}{*}{$8.1,8.7$} & \multirow{3}{*}{$5.2{\times}10^4$} & $246.0$ & $782.1$ & $1372.6$ & $1733.0$ & $1.42$ & $0.166$ & $1.45$ & $0.165$ \\
 & & & & $184.6$ & $782.1$ & $1386.7$ & $1799.7$ & $1.31$ & $0.163$ & $1.34$ & $0.162$ \\
 & & & & $147.7$ & $782.1$ & $1404.3$ & $1799.8$ & $1.45$ & $0.161$ & $1.50$ & $0.160$ \\
\addlinespace
\multirow{3}{*}{SS1} & \multirow{3}{*}{$0.5{+}0.5$} & \multirow{3}{*}{$8.4$} & \multirow{3}{*}{$5.2{\times}10^4$} & $221.5$ & $806.8$ & $1431.7$ & $1774.9$ & $0.40$ & $0.147$ & $0.45$ & $0.142$ \\
 & & & & $166.1$ & $796.5$ & $1438.6$ & $1724.9$ & $1.27$ & $0.152$ & $1.38$ & $0.148$ \\
 & & & & $132.9$ & $825.0$ & $1447.0$ & $1743.0$ & $1.14$ & $0.147$ & $1.25$ & $0.143$ \\
\addlinespace
\multirow{3}{*}{SS1}  & \multirow{3}{*}{$0.4{+}0.6$}   & \multirow{3}{*}{$7.8,8.9$}   & \multirow{3}{*}{$5.2{\times}10^4$} & $184.6$  & $795.2$ & $1281.1$ & $2066.4$ & $2.03$ & $0.160$ & $2.09$ & $0.158$ \\
 & & & & $147.7$ & $804.9$ & $1296.7$ & $2066.4$ & $1.88$ & $0.163$ & $1.93$ & $0.161$ \\
 & & & & $123.0$ & $818.2$ & $1300.3$ & $2032.2$ & $1.99$ & $0.161$ & $2.05$ & $0.159$ \\
\addlinespace
\multirow{3}{*}{SS1} & \multirow{3}{*}{$0.7{+}0.7$} & \multirow{3}{*}{$9.3$} & \multirow{3}{*}{$1.4{\times}10^4$} & $246.1$ & $827.6$ & $1482.3$ & $1866.4$ & $0.54$ & $0.174$ & $0.58$ & $0.171$ \\
 & & & & $184.6$ & $827.6$ & $1502.1$ & $1866.4$ & $1.19$ & $0.168$ & $1.26$ & $0.166$ \\
 & & & & $147.7$ & $807.7$ & $1509.5$ & $1891.6$ & $1.21$ & $0.167$ & $1.27$ & $0.164$ \\
\addlinespace
\multirow{3}{*}{SS2} & \multirow{3}{*}{$0.5{+}0.5$} & \multirow{3}{*}{$8.0$} & \multirow{3}{*}{$4.0{\times}10^4$} & $204.3$ & $865.2$ & $1522.9$ & $1916.4$ & $1.02$ & $0.154$ & $1.10$ & $0.151$ \\
 & & & & $153.2$ & $868.7$ & $1569.3$ & $1853.7$ & $1.24$ & $0.150$ & $1.35$ & $0.147$ \\
 & & & & $122.6$ & $865.2$ & $1566.4$ & $1916.6$ & $0.76$ & $0.144$ & $0.85$ & $0.140$ \\
\addlinespace
\multirow{3}{*}{SS3} & \multirow{3}{*}{$0.5{+}0.5$} & \multirow{3}{*}{$7.7$} & \multirow{3}{*}{$3.2{\times}10^4$} & $196.9$ & $910.8$ & $1649.0$ & $1999.8$ & $0.79$ & $0.160$ & $0.85$ & $0.156$ \\
 & & & & $147.7$ & $920.2$ & $1678.0$ & $1999.7$ & $1.06$ & $0.156$ & $1.14$ & $0.153$ \\
 & & & & $118.1$ & $928.5$ & $1664.6$ & $1999.9$ & $1.04$ & $0.148$ & $1.10$ & $0.146$ \\
\midrule
\multirow{3}{*}{SFHo} & \multirow{3}{*}{$0.4{+}0.4$} & \multirow{3}{*}{$12.7$} & \multirow{3}{*}{$2.4{\times}10^5$} & $270.7$ & $570.0$ & $941.5$ & $1699.8$ & $1.12$ & $0.129$ & $1.03$ & $0.132$ \\
 & & & & $203.0$ & $592.1$ & $1042.2$ & $1749.9$ & $1.00$ & $0.110$ & $0.88$ & $0.113$ \\
 & & & & $162.4$ & $601.1$ & $1048.4$ & $1784.1$ & $1.15$ & $0.114$ & $1.03$ & $0.116$ \\
\addlinespace
\multirow{3}{*}{SFHo} & \multirow{3}{*}{$0.45{+}0.55$} & \multirow{3}{*}{$12.5,12.3$} & \multirow{3}{*}{$8.3{\times}10^4$} & $246.1$ & $647.5$ & $989.3$ & $1883.8$ & $1.14$ & $0.150$ & $1.05$ & $0.154$ \\
 & & & & $184.6$ & $668.0$ & $1014.1$ & $1949.9$ & $1.05$ & $0.148$ & $0.97$ & $0.152$ \\
 & & & & $147.7$ & $666.2$ & $1017.4$ & $1849.8$ & $0.83$ & $0.153$ & $0.76$ & $0.157$ \\
\addlinespace
\multirow{3}{*}{SFHo} & \multirow{3}{*}{$0.5{+}0.5$} & \multirow{3}{*}{$12.4$} & \multirow{3}{*}{$8.2{\times}10^4$} & $246.1$ & $663.3$ & $1141.8$ & $1933.1$ & $0.81$ & $0.124$ & $0.72$ & $0.128$ \\
 & & & & $184.6$ & $675.8$ & $1176.2$ & $1888.7$ & $0.90$ & $0.133$ & $0.82$ & $0.136$ \\
 & & & & $147.7$ & $681.0$ & $1183.0$ & $1909.0$ & $1.00$ & $0.137$ & $0.93$ & $0.139$ \\
\addlinespace
\multirow{3}{*}{SFHo} & \multirow{3}{*}{$0.4{+}0.6$} & \multirow{3}{*}{$12.7,12.2$} & \multirow{3}{*}{$8.7{\times}10^4$} & $246.1$ & $636.4$ & $820.3$ & $1799.8$ & $1.94$ & $0.169$ & $1.86$ & $0.171$ \\
 & & & & $184.6$ & $643.1$ & $834.1$ & $1799.8$ & $1.80$ & $0.171$ & $1.73$ & $0.173$ \\
 & & & & $147.7$ & $624.0$ & $818.3$ & $1737.0$ & $1.78$ & $0.173$ & $1.71$ & $0.175$ \\
\addlinespace
\multirow{3}{*}{SFHo} & \multirow{3}{*}{$0.7{+}0.7$} & \multirow{3}{*}{$12.1$} & \multirow{3}{*}{$1.6{\times}10^4$} & $184.6$ & $782.2$ & $1393.6$ & $2066.4$ & $0.53$ & $0.145$ & $0.47$ & $0.149$ \\
 & & & & $147.7$ & $794.3$ & $1418.3$ & $2219.1$ & $0.40$ & $0.145$ & $0.35$ & $0.151$ \\
 & & & & $123.0$ & $797.0$ & $1417.6$ & $2133.3$ & $0.60$ & $0.151$ & $0.55$ & $0.155$ \\
\addlinespace
\multirow{3}{*}{WFF1} & \multirow{3}{*}{$0.5{+}0.5$} & \multirow{3}{*}{$10.4$} & \multirow{3}{*}{$3.3{\times}10^4$} & $246.1$ & $849.3$ & $1490.1$ & $2266.4$ & $0.23$ & $0.112$ & $0.32$ & $0.106$ \\
 & & & & $184.6$ & $872.5$ & $1595.9$ & $2346.5$ & $0.22$ & $0.116$ & $0.29$ & $0.110$ \\
 & & & & $147.7$ & $881.6$ & $1620.9$ & $2333.0$ & $0.60$ & $0.131$ & $0.66$ & $0.128$ \\
\addlinespace
\multirow{3}{*}{DD2} & \multirow{3}{*}{$0.5{+}0.5$} & \multirow{3}{*}{$13.1$} & \multirow{3}{*}{$1.1{\times}10^5$} & $270.7$ & $630.0$ & $1077.3$ & $1733.0$ & $1.11$ & $0.122$ & $1.02$ & $0.124$ \\
 & & & & $203.0$ & $619.5$ & $1107.8$ & $1666.5$ & $1.35$ & $0.131$ & $1.26$ & $0.134$ \\
 & & & & $162.4$ & $619.5$ & $1124.9$ & $1767.1$ & $1.01$ & $0.134$ & $0.91$ & $0.137$ \\
\bottomrule
\end{tabular}
\end{table}

\subsection{Code performance of \sacrak}
\label{app:performance}

The simulations presented in this work are performed with \sacrak on the
Viper APU cluster of the Max Planck Computing and Data
Facility\footnote{\url{https://docs.mpcdf.mpg.de/doc/computing/viper-gpu-user-guide.html}}.
\sacrak uses a nested Cartesian grid with $2{:}1$ box-in-box adaptive mesh
refinement and mirror symmetry about the orbital ($z=0$) plane. The grid has
ten refinement levels: each star is covered by four finer comoving domains, while
six coarser domains enclose both stars and are centered on the binary's center of
mass. Each domain is a uniform cell-centered Cartesian grid of
$2N{\times}2N{\times}N$ cells in $(x,y,z)$, with six buffer cells on each side. The
grid spacing on level $\ell$ is $\Delta x^{(\ell)}=L^{(\ell)}/(2N)$, where
$L^{(\ell)}$ is the size of the domain in the $x$ and $y$ directions. Every level is
split into $\mathtt{jprocs}{\times}\mathtt{kprocs}{\times}\mathtt{lprocs}$ subdomains
in $(x,y,z)$, and each MPI rank evolves one subdomain across all levels, exchanging
boundary data through GPU-aware MPI. \Cref{tab:performance} lists the setup and
measured performance for a representative binary, whose three resolutions $N=60$,
$80$, and $100$ use $1$, $2$, and $4$ nodes at $\approx3\,\mathrm{s}$ per step and
advance $50$--$106\,\mathrm{ms}$ of physical evolution per day.

\begin{table}[!ht]
\caption{Parallel setup, per-step wall-clock time, and physical evolution
covered per day for \sacrak on the Viper APU nodes, for the
representative $0.5{+}0.5\,M_\odot$ SFHo binary at its three grid resolutions.}
\label{tab:performance}
\begin{tabular}{c c c c c c c}
\toprule
$N$ & $\Delta x_\mathrm{finest}$ [m] & nodes & $\mathtt{jprocs}{\times}\mathtt{kprocs}{\times}\mathtt{lprocs}$ & ranks & $t_\mathrm{step}$ [s] & $t_\mathrm{phys}$/day [ms] \\
\midrule
$60$  & $246.1$ & $1$ & $2{\times}2{\times}1$ & $4$ & $2.7$ & $106$ \\
$80$  & $184.6$ & $2$ & $2{\times}2{\times}1$ & $4$ & $3.3$ & $64$ \\
$100$ & $147.7$ & $4$ & $4{\times}2{\times}1$ & $8$ & $3.4$ & $50$ \\
\bottomrule
\end{tabular}
\end{table}

\end{document}